\documentclass[11pt]{article}
\usepackage{amsmath,amssymb,amscd}
\usepackage[usenames]{color}
\usepackage{graphicx}
\usepackage{epstopdf}
\usepackage[mathscr]{eucal}
\usepackage{lineno}
\usepackage{color}
\usepackage{soul}
\usepackage{cite}
\usepackage{subfigure}
\usepackage{authblk}

\large\normalsize

\begin{document}
\title{Exponential Cell Division and Allometric Scaling in Metabolic Ecology}

\author[ab1]{Jia-Xu Han}
\author[a1]{Zhuangdong Bai}

\author[a*]{Rui-Wu Wang}

\affil[a]{School of Ecology and Environment, Northwestern Polytechnical University, Xi'an 710072, PR China}
\affil[b]{Zoology Department and Biodiversity Research Centre, University of British Columbia, Vancouver, British Columbia V6T 1Z4, Canada}
\affil[*]{Corresponding author: Rui-Wu Wang (wangrw@nwpu.edu.cn)}

\affil[1]{These authors contributed equally to this work.}

\date{}

 \maketitle
\newpage

\begin{center}
\section*{Abstract}

\end{center}
One of the most fundamental rules in metabolic ecology is the allometric equation, which is a power-law scaling that describes the connection between body measurements and body size. The biological dynamics of this essentially empirical allometric equation, however, have yet to be properly addressed in cell level. In order to fill the gap between biological process in cell level and allometric scaling in metabolic ecology, we simply assumed a cell bipartition without limitation, and then exponential cells increased during their lifetime. Two synchronous exponential increasing could generate a power-law scaling between body mass and an organ's weight. And the power-law scaling between body mass and metabolic rate may also be obtained by substituting an organ's weight with the weight of erythrocytes. Based on the same assumption, the dynamic of cell proliferation reveal a complex exponential scaling between body mass and longevity rather than the previously reported power-law scaling. In other words, there is a quadratic relationship between longevity and logarithmic form of body mass. In these relationships, all parameters can be explained by indices in cell division and embryo.

\textbf{key words:}Allometric Equation, Longevity, Body Size, Metabolic Rate.

\newpage

\section{Introduction}
In the late 19th century, the correlation between body size and other measurements was initially identified through the utilization of a log-log plot, which demonstrated the direct relationship between anthropoid brain weight and body weight\cite{stahl1965}. Following that, hundreds of experimental and theoretical studies on the relationship between body weight and other body measures, including organ weight, body fluids, lifespan, and metabolic rate, were conducted to demonstrate the relevance of body size in biological lives\cite{brody1945}\cite{stahl1962}\cite{calder1983}\cite{schmidt1984}\cite{donabedian2001}. Allometric equations, which are power-law scaling, are used to explain the majority of body measures in relation to body size\cite{huxley1932}. And it is generally expressed as
\begin{eqnarray}
Y=Y_0M^b, \nonumber
\end{eqnarray}
where $Y$ is a body measurement such as longevity, organ weight, or metabolic rate, $M$ is usually body mass, $Y_0$ is a normalization constant, and $b$ is an allometric exponent\cite{brown2004}. The relationship between metabolic rate and body mass is the most well-known and contentious of the size-dependent issues. The metabolic theory of ecology demonstrated the $\frac{3}{4}$-power rule fulfilled for a range of over 21 orders of magnitude from bacteria to whales and ushered in a new era of research, although there are still questions\cite{west1997}\cite{savage2008}\cite{isaac2010}. Furthermore, providing the concept of the fourth spatial dimension generated from fractal-like networks gives a framework of quarter-power scaling, into which most body measures are incorporated. However, all of these efforts, as well as some other models, are based on the premise that there is a directly proportionate link between metabolic rates and power-law scaling of body mass, which is not supported by theoretical evidence\cite{west1999}\cite{kozlowski2003}\cite{speakman2010}\cite{glazier2010}. In addition,all of these studies are primarily concerned with the value of the allometric exponent while ignoring the normalization constant. 

Scaling laws should somehow reflect processes at lower levels such as cell division. Cell division results in the dynamic of cell number corresponding to body mass\cite{kozlowski2003}. Here, we consider the development of relationships between body size and other measures in hypothetical lineages of species descended from similar-sized ancestor species. The word lineage refers to a collection of closely related species, living or extinct, that have the same pattern of adult body size change in relation to changes in cell size and cell quantity. In this context, we analyze the dynamic of cell number with appropriate resources in order to demonstrate the relationship between body mass and various body measures. In section 2, we merge two synchronous exponential dynamics to give an explanation of power-law scaling between body mass and body measurements except longevity. Meanwhile, we can give a microscopic explanation about parameters, including the ignored normalization constant, in this power-low scaling. In section 3, we use cell division dynamics to build the relationship between body mass and longevity. To determine the appropriate model, we evaluate the (complex) exponential scaling against power-law scaling, using experimental data that has been reported. Finally, we give a discussion in section 4.

\section{The relationship between the individual and its any organ}
Our model starts with a larva or embryo that has had its different tissues separated\cite{hayflick1962}. Although the initial condition is a fuzzy time, it has no effect on the model's findings (see details in Appendix A). Although cells vary in size, cell cycle, and death rate among tissues, we use the 'average' cell size, cell cycle, and mortality rate in subsequent considerations\cite{kozlowski2003}.This individual's life might be split into segments in which the number of cells proliferates fixed multiple, and each segment is designated as a generation. There are hundreds of distinct cell kinds, each with its own cell circle. During a cell cycle, each cell splits into two. Each generation is the average result of various types of cell cycles. We assume that each segment has the same duration. At generation $t$, the dynamic of the individual and its any organ's cell number (or two organ's cell number) may be expressed as 
\begin{eqnarray}
n=n_1r_1^t ,   \\
n'=n_2r_2^t,
\end{eqnarray}
where $n_1$ and $n_2$ represent the individual's and its any organ's (or two organ's) starting number of cells at time $t=0$, and $r_1$ and $r_2$ represent the individual's and its any organ's (or two organ's) cell proliferation rate. After calculating the logarithm of these two equations (Eqs.(1) and (2)), we may get 
\begin{eqnarray}
\ln n'=\frac{\ln r_2}{\ln r_1}\ln n+\ln n_2-\frac{\ln r_2}{\ln r_1}\ln n_1.
\end{eqnarray}
And the weight of the individual and any organ (or two organs) is expressed as 
\begin{eqnarray}
w=nw_1 ,   \\
w'=n'w_2 ,
\end{eqnarray}
where $w_1$ and $w_2$ are the average cell mass of the individual and its any organ (or two organs). The relationship between an individual's weight and the weight of any organ (or two organs) is  
\begin{eqnarray}
\ln w'=\frac{\ln r_2}{\ln r_1}\ln w+\ln (n_2w_2)-\frac{\ln r_2}{\ln r_1}\ln (n_1w_1).
\end{eqnarray}
Species from the same ancestry often have comparable cell proliferation processes and organ kinds. These species should have the same parameter in Eq.(6), and there should be a linear relationship between the logarithmic form of the individual's and its any organ's (or two organ's) weight. This relationship is a power-law scaling induced by the individual's and any organ's distinct cell proliferation rate. The slope of a linear relationship is defined by the ratio of the individual's logarithmic cell proliferation rate to that of any organ (or two organs). The intercept of this linear relationship is influenced by the ratio of the individual's and its any organ's (or two organ's) logarithmic form of cell proliferation rate, as well as the initial condition of the individual's and its any organ's (or two organ's) weight. This part of the conclusion is compatible with Huxley's work\cite{huxley1932}, and we utilize cell division to demonstrate the varied growth rates in order to obtain additional information about parameters. 

In the experiment, metabolic rate is generally determined by the amount of oxygen or energy expended per unit weight per unit time \cite{lasiewski1971}. Geoffrey B. West utilized a positive correlation between metabolic rate and fluid flow velocity in his mathematical model because fluid carries oxygen and nutrients for metabolism\cite{west1997}. Here, we look at how cells or fluids transport oxygen and nutrients, using erythrocytes as an example. Because erythrocytes carry oxygen for metabolism, the quantity of erythrocytes should be proportional to metabolic rate. Although erythrocytes do not proliferate on their own, they are generated from erythroid progenitor cells, which can proliferate in the same way as organ cells do. The erythrocyte dynamic is thus approximated to that of organ cells by disregarding the complicated cell development from erythroid progenitor cells to erythrocytes. To summarize, as previously established, there should be a linear relationship between logarithmic form of body mass and logarithmic form of metabolic rate. By viewing erythrocytes as an organ, the genesis of power-law scaling between metabolic rate and body mass becomes obvious.Coincidentally, this power-law scaling appears to be such a general norm that it is applicable to all natural systems, from molecules and mitochondria to cells and humans\cite{west2002}. We believe this is due to the same micro process of living cells. On average, species from various lineages exhibit very minor differences in cell proliferation. Fluids might be considered as the creation of certain cells and in direct proportion to the quantity of cells. The power-law scaling between fluids and body mass will then be compatible with experimental data patterns\cite{stahl1962}$^,$\cite{stahl1963}. And such power-law scaling have been test broadly in experiments\cite{west1997}.

\section{The relationship between body mass and longevity}
\subsection{Constant growth rate}
Similarly, we starts with a larva or embryo. In this starting state, the individual has $n_0$ cells and the generation is $t=0$. The body mass ($w$) of an individual can be estimated as 
\begin{eqnarray}
w=nw_0,
\end{eqnarray}
where $n$ represents the number of cells and $w_0$ represents the average cell mass. Normal cells lose telomeres with each cell division, and short telomeres are ineffective in protecting cells from DNA damage and cellular senescence\cite{blasco2003}\cite{shay2011}. Cells with a limited ability to divide offer a method to quantify longevity. And this individual's longevity ($L$) may be expressed as 
\begin{eqnarray}
L=Tl_0,
\end{eqnarray}
where $T$ is the number of generations and $l_0$ is the duration between generations. The proliferation is exponential, and the relationship between the number of cells and generations may be expressed as 
\begin{eqnarray}
n=n_02^T.
\end{eqnarray}
We may derive the connection between longevity and body mass from these equations (Eqs. (7), (8), and (9))
\begin{eqnarray}
\ln w=\frac{\ln 2}{l_0}L+\ln (w_0n_0).
\end{eqnarray}
Similarly, these species should have the same parameter in Eq.(10), and there should be a linear relationship between longevity and logarithmic body mass for all members of the lineage. It is worth mentioning that the relationship between longevity and body mass exhibits exponential increase rather than the previously reported power-law growth. The time spent in each generation determines the slope of the linear relationship. Furthermore, the intercept of the linear relationship is influenced by the initial condition of the larva's weight. This concept might potentially be applied to the link between organ weight and longevity. We can also obtain a linear relationship between longevity and the logarithmic form of organ weight. 

\subsection{Constant growth rate and cell death}
In section 3.1, the assumption that cells never die is critical for exponential increase. It can only occur in certain circumstances, such as juvenile animals and microorganisms\cite{sibly1986}\cite{sibly2020}\cite{PAULTON1991}\cite{Zwietering1990}. Furthermore, we consider replicative aging, which is a more realistic scenario that cell mortality increases over time\cite{Ackermann2003}. The dynamic of cells number can be written as
\begin{eqnarray}
&&n_{i+1}=2a(1-kt)n_i, \nonumber \\ 
&&i=0,1,2,\cdots,T-1,
\end{eqnarray}
where $n_i$ is cells number at generation $t=i$, $a$ is the mortality rate at initial condition, and $k$ is the increasing rate in mortality per generation. We can get
\begin{eqnarray}
\ln n_{T}=\ln n_0+T\ln (ar)+\sum\limits_{i=0}^{T-1}\ln (1-ki). 
\end{eqnarray}
If $k<<1$, this equation can be approximately written as
\begin{eqnarray}
\ln n_{T}=\ln n_0+T\ln (ar)-\sum\limits_{i=0}^{T-1}ki. 
\end{eqnarray}
Eq.(7) and Eq.(8) can be written as
\begin{eqnarray}
&&w=n_Tw_0,  \\
&&L=Tl_0. 
\end{eqnarray}
And we can get a quadratic function between longevity and logarithmic form of body mass written as
\begin{eqnarray}
\ln w=-\frac{k}{2l_0^2}L^2+\frac{\frac{k}{2}+\ln (2a)}{l_0}L+\ln (n_0w_0). 
\end{eqnarray}
Here, the logarithmic form of body mass is a quadratic function of longevity.

\subsection{Model selection}

To compare our models with the power-law scaling between body mass and longevity, we utilize a reported data to perform empirical studies of the link between longevity and body mass\cite{Tacutu2012}. After removing partial data, the assembled data contains 530 species from four classes:  \emph{Aves} (167 species), \emph{Mammalia} (349 species), and \emph{Reptilia} (14 species). Figure 1 depicts the relationship between the logarithmic form of body mass and longevity for all 530 species. It appears that various classes are distributed differently. To highlight trends in these data, we test the quadratic and linear relationship in each class separately in Figure 2. And, we also show the linear relationship between the logarithmic form of body mass and logarithmic form of longevity. 

\begin{figure}[htb]
\includegraphics[width=\textwidth]{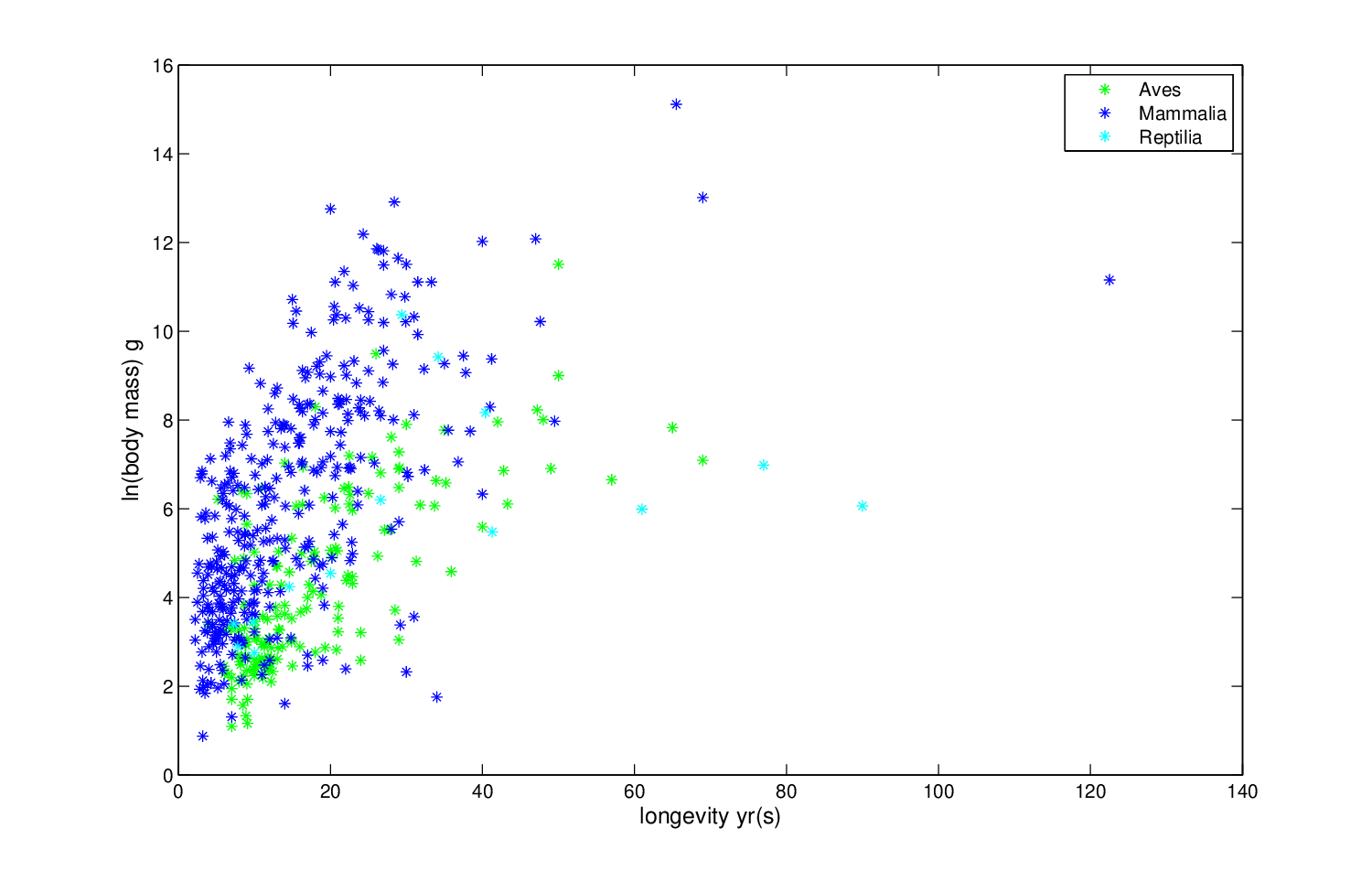}
\caption{The relationship between body mass in logarithmic form and longevity. All 530 species in four classes are shown in various colors, and the distribution of each class varies. }
\label{fig1}
\end{figure}

\begin{figure}[htb]
\centering
\subfigure[Aves]{
\begin{minipage}[b]{0.3\textwidth}
\includegraphics[width=\textwidth]{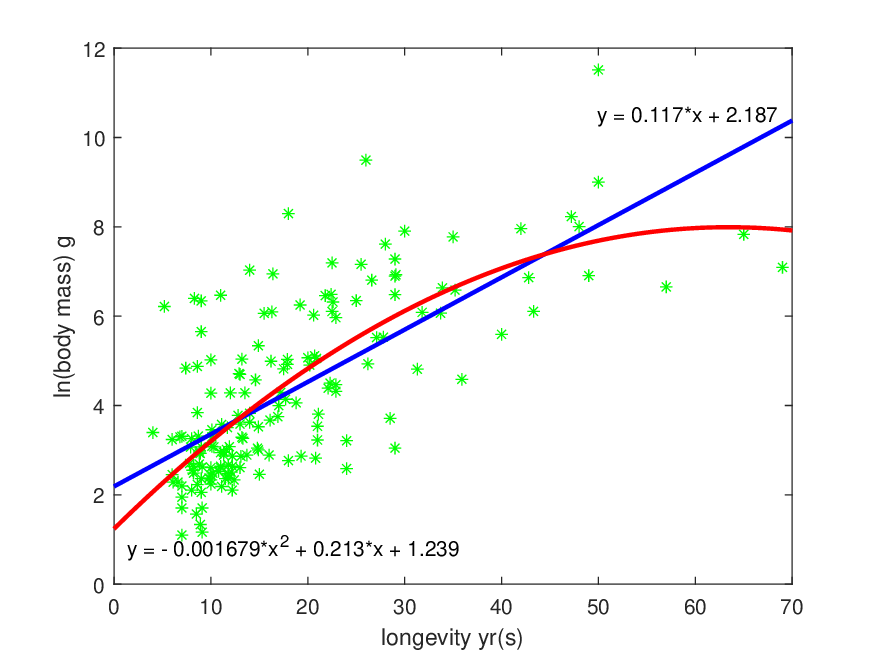} \\
\includegraphics[width=\textwidth]{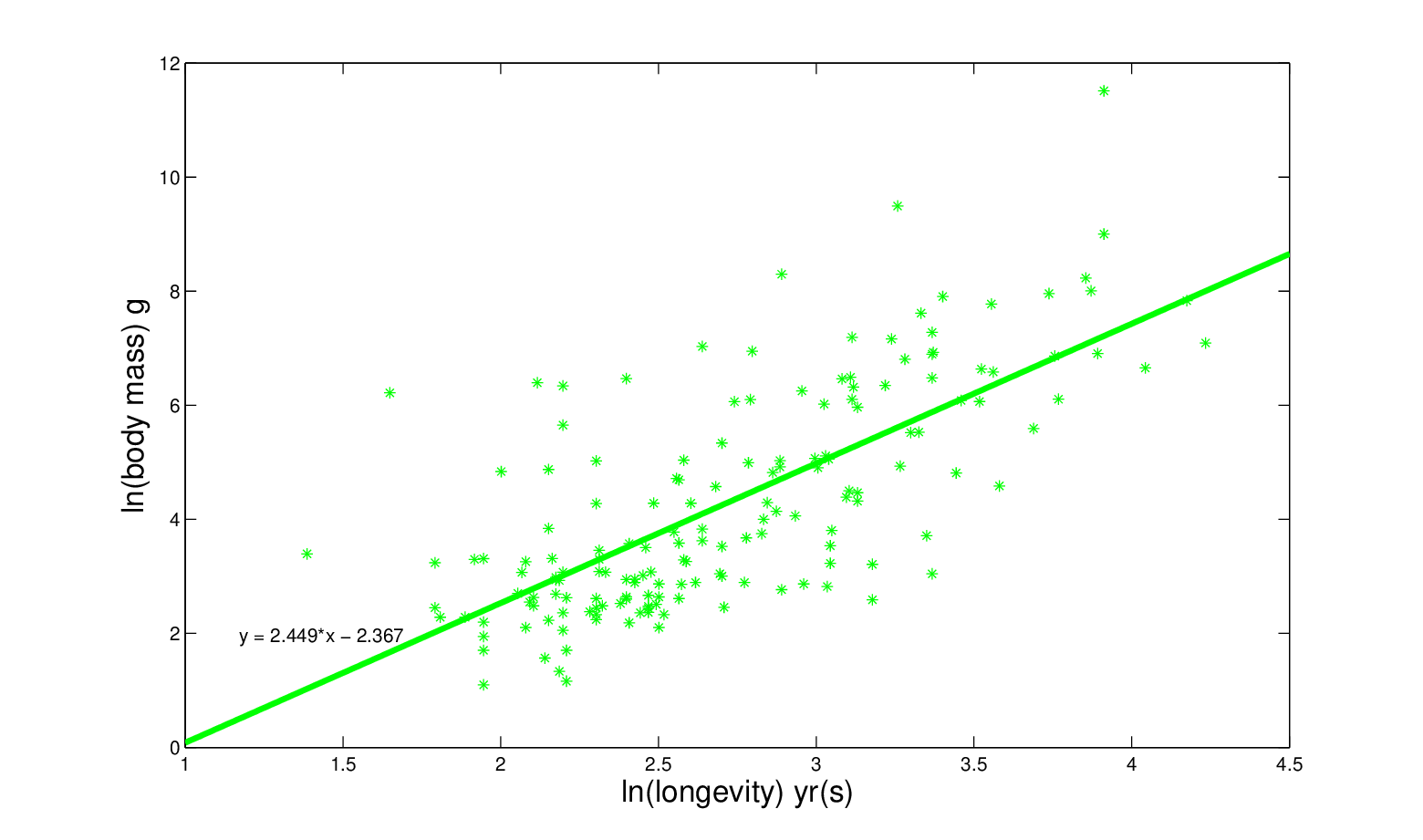}
\end{minipage}
}
\subfigure[Mammalia]{
\begin{minipage}[b]{0.3\textwidth}
\includegraphics[width=\textwidth]{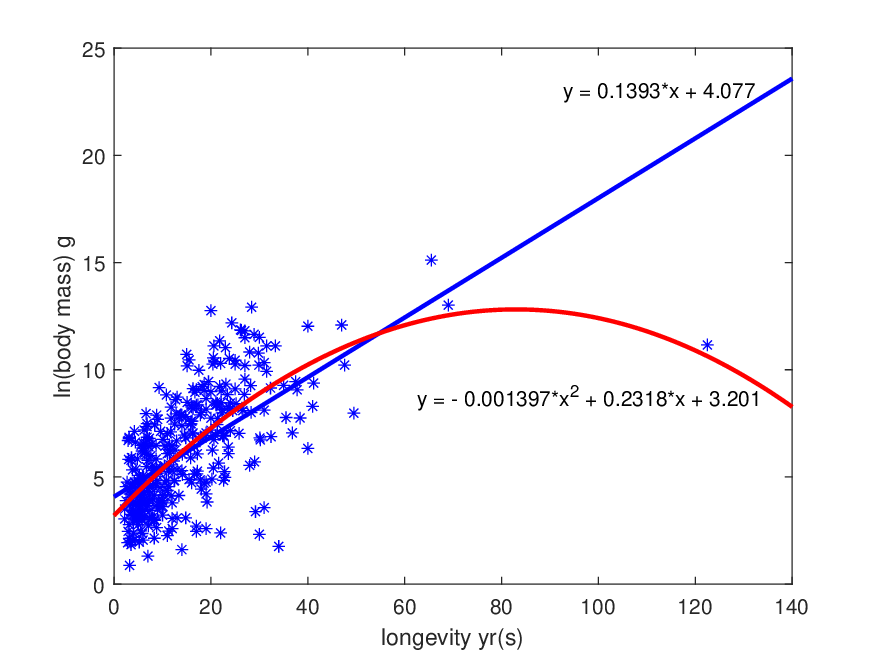} \\
\includegraphics[width=\textwidth]{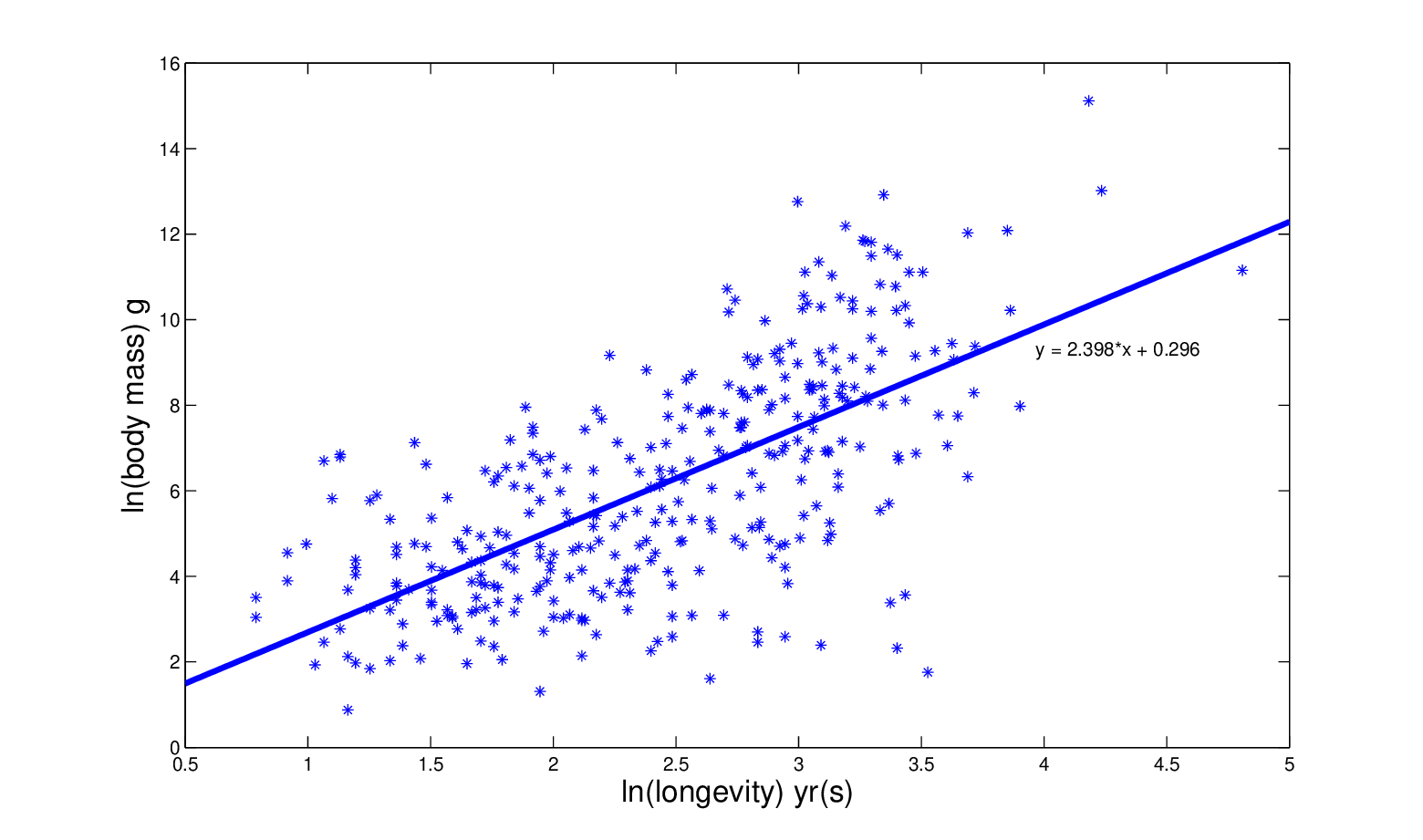}
\end{minipage}
}
\subfigure[Reptilia]{
\begin{minipage}[b]{0.3\textwidth}
\includegraphics[width=\textwidth]{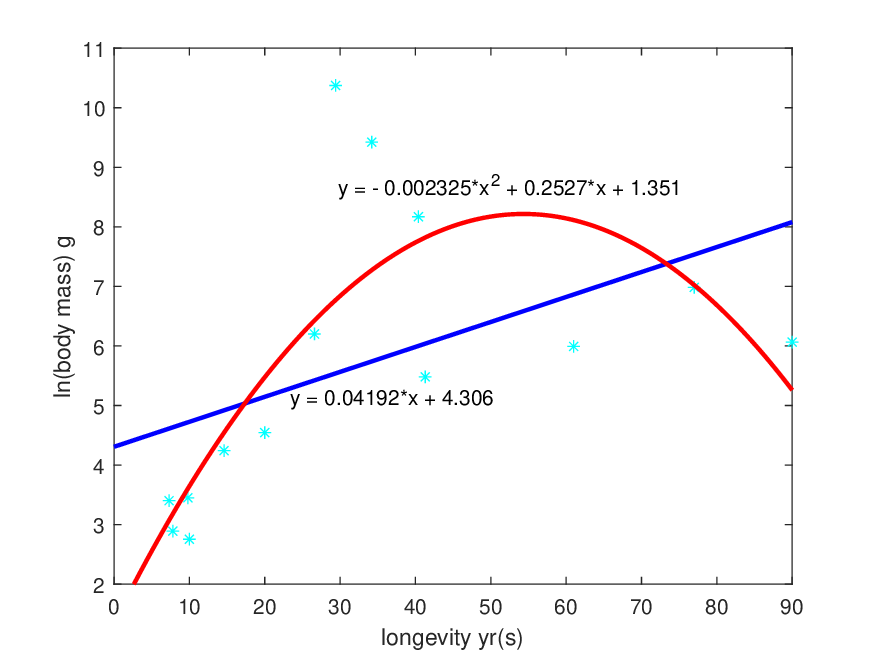} \\
\includegraphics[width=\textwidth]{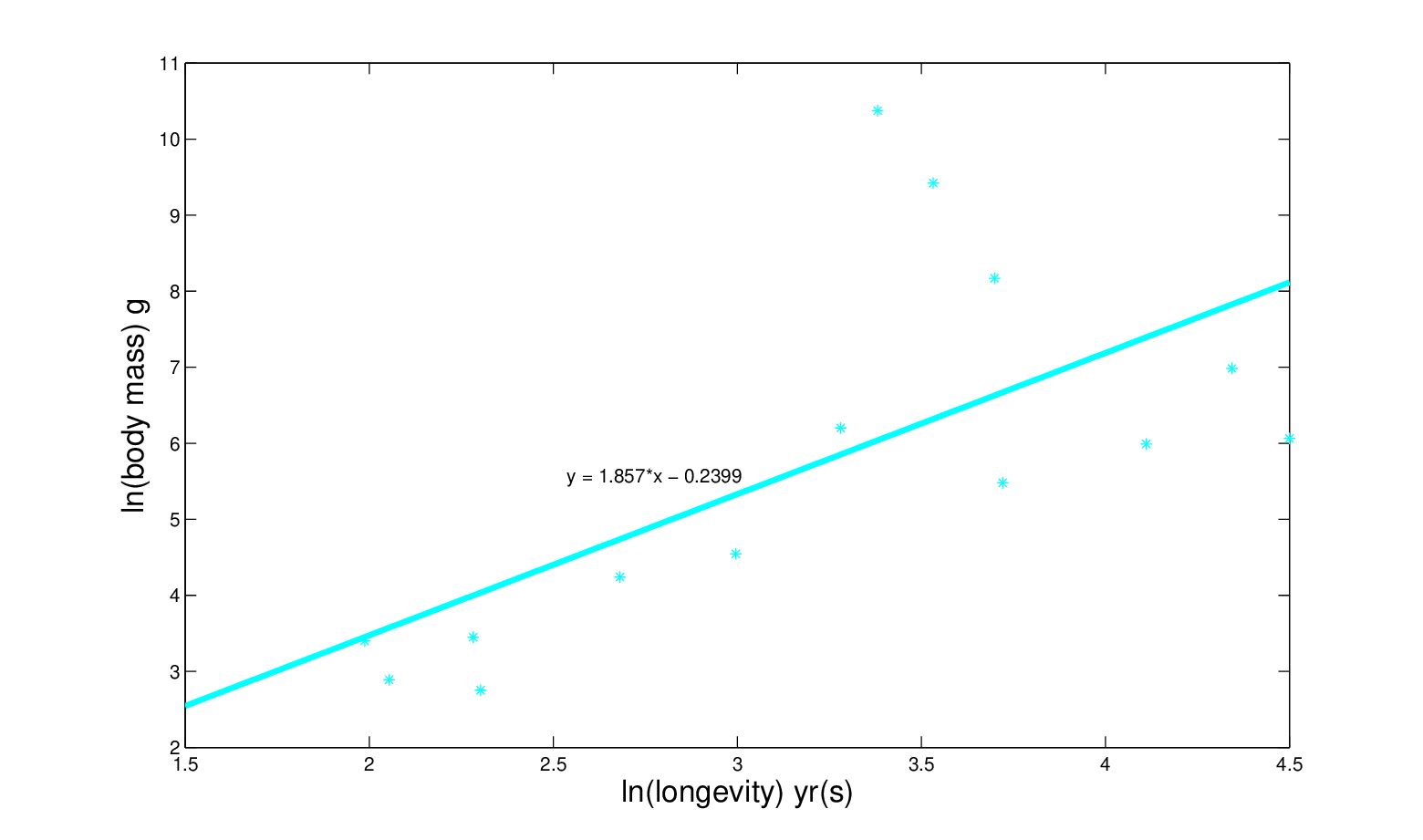}
\end{minipage}
}
\caption{The relationship between body mass and longevity across several lineages. Three classes, \textbf{a.} \emph{Aves}, \textbf{b.} \emph{Mammalia} and \textbf{c.} \emph{Reptilia} , are plotted respectively. The first row show the relationship between logarithmic form of body mass and longevity with linear and quadratic regression. The second row show the relationship between logarithmic form of body mass and logarithmic form of longevity with linear regression.}
\label{fig2}
\end{figure}

Here, Akaike information criterion (AIC) and Mallow's $C_p$ (M$C_p$) are used for model selection\cite{Burnham2004}\cite{Kobayashi1990}\cite{Mallows2012}. All results of statistical analysis are provided in Table 1. In \emph{Aves}, \emph{Mammalia} and \emph{Reptilia}, the quadratic relationship between logarithmic form of body mass and longevity does better than the other two models because it has lower SSE. And because the AIC and M$C_p$ of quadratic relationship are lower than the others in these classes, the quadratic relationship between logarithmic form of body mass and longevity is selected to describe the relationship between body mass and longevity. It also means that the relationship between body mass and longevity is a complex exponential form other than power-law scaling. It's worth noting that the parameters $a$ in quadratic relationship are negative in \emph{Aves}, \emph{Mammalia} and \emph{Reptilia}, which is consistent with Eqs.(16).

\begin{table}[htb]
\caption{Different models between body mass ($w$) and longevity ($L$).}
\begin{tabular}{c|ccccccc}
model & Class    & a         & b       & c     & SSE         & AIC     & M$C_p$    \\ \hline
$ln(w)=a ln(L)+b$     & Aves     & 2.449     & -2.367  &       & 308.6   & 3.4759  &1.9155 \\
      & Mammalia & 2.398     & 0.296   &       & 1379    & 4.2234  &4.0198 \\
      & Reptilia & 1.857     & -0.2399 &       & 42.49 & 4.2338 &4.6905 \\ \hline
$ln(w)=a L+b$     & Aves     & 0.117     & 2.187   &       & 310.2  & 3.4811 &1.9254 \\
      & Mammalia & 0.1393    & 4.077   &       & 1514    & 4.3168 &4.4133 \\
      & Reptilia & 0.04192   & 4.306   &       & 58.51  & 4.5537 &6.4589 \\  \hline
$ln(w)=a L^2+bL+c$     & Aves     & -0.001679 & 0.213   & 1.239 & 289     & 3.4103 &1.8155 \\
      & Mammalia & -0.001397 & 0.2318  & 3.201 & 1344    & 4.1977 &3.9403 \\
      & Reptilia & -0.002325 & 0.2527  & 1.351 & 30.41   & 3.8993 &3.9099
 
\end{tabular}
\end{table}

\section{Discussion}
In nature, power-law scaling is prevalent. It is critical to comprehend how it is formed as well as the significance of its characteristics. The dynamic of cell number was modeled in this article to develop the link between body size and many other aspects of biology from micro level. We discover an essential approach to explain the formation of power-law scaling here. We can observe that two synchronous exponential growth processes can result in the occurrence of the power-law scaling relationship. Although the exponential increase can only occur in perfect situations, the notion in this work may also be utilized to analyze non-linear phenomena in cancer or bacterium proliferation, particularly for some exponential or power-law patterns. 

There is a lengthy debate in metabolic ecology regarding the scaling exponent in order to identify a universal rule in nature, however the normalization constant is frequently overlooked\cite{isaac2010}. Our model demonstrates that the initial condition has a significant influence on the normalization constant. Because the tissues have been divided in this early stage, cell differentiation has minimal effect. It appears that the manner of producing will have an effect on the normalization constant. Alternatively, we may argue that this normalization can be used to categorize species, and that this categorization is based on the mode of producing. 

A special body measurement is longevity. The similar cell proliferation in animals descended from the same ancestor is responsible for the connection between longevity and body mass. Based on the dataset analysis, this cell growth appears to be restricted to a narrow region across animal groups. It is likely due to homology in evolution, which is also one of our model's assumptions, that species with similar ancestors have comparable cell proliferation processes. Based on our model, difference of body mass among species from the same ancestry is produced by different times of cell division in the same growth curve. In this context, we find that the quadratic relationship between logarithmic form of body mass and longevity has better performance than the power-law scaling between body mass and longevity. It's worth noting that the quadratic relationship in Eqs.(16) has negative quadratic coefficient. It means that there is a maximum body size even some species have very long life but there is not in power-law scaling. In nature, there are plenty of evidences on the limits of body size\cite{Marquet1998}\cite{Smith2011}.

Overall, the main goal of this model is to provide a mechanistic explanation of allometric relationships and provide a better model to explain the relationship between body mass and longevity than before. Our model demonstrates the origins of allometric equations such as a complex exponential scaling between body mass and longevity and power-law scaling between body mass and some body measurements. The most significant explanation for these results is cell division, which is a natural event in micro level. This model is a little simple and ignores the fact that during differentiation several cells may decrease their division rates and, in extreme cases such as in neurons, exhibit extremely low division rates under very particular circumstances. So, extending our model to other empirical growth curves should be considered in the future.

\section*{Authorship contribution statement}
\textbf{Jia-Xu Han:} Conceptualization, Methodology, Investigation, Writing-Original Draft, Visualization.  \textbf{Zhuangdong Bai:} Conceptualization, Formal analysis, Investigation,  Writing-Original Draft, Visualization. \textbf{Rui-Wu Wang:} Conceptualization,  Writing-Original Draft, Supervision, Project administration, Funding acquisition, Visualization.
\section*{Declaration of Competing Interest}
The authors declare that they have no known competing financial interests or personal relationships that could have appeared to influence the work reported in this paper.

\section*{Acknowledgments}
We thank Cong Li for many useful suggestions that improved the quality of this article. This research was supported by National Natural Science Foundation of China-Yunnan Joint Fund (U2102221).

\section*{Appendix A: Initial condition does not affect the result}
We adopt a different initial condition in the relationship between lifespan and body mass, starting a bit later, and the equations may be expressed as 
\begin{eqnarray}
&&w=nw_0, \nonumber \\ \nonumber
&&L'=T'l_0, \\ \nonumber
&&n=n'_02^{T'}, \\ \nonumber
&&T'=T-t, \\ \nonumber
&&L'=L-tl_0, \\ \nonumber
&&n'_0=n_02^t,
\end{eqnarray}
where $L'$, $T'$, and $n'_0$ have the same definition as $L$, $T$, and $n_0$ but with a new starting condition and $t_0$ is the time lag between two initial conditions. We are able to obtain 
\begin{eqnarray}
L'=\frac{l_0}{\ln 2}\ln w-\frac{l_0}{\ln 2}\ln (w_0n'_0). \nonumber 
\end{eqnarray} 
from the first three equations and the last two equations make this equation be equivalent to Eq.(4).

In the relation between the weights of two organs, we also employ a new initial condition in which we start a bit later, with a time lag of $t_0$.
Equations can be written as follows:  
\begin{eqnarray}
&&N=N_1r_1^t, \nonumber \\ \nonumber
&&N'=N_2r_2^t, \\ \nonumber
&&W=Nw_1, \\ \nonumber
&&W'=N'w_2, \\ \nonumber
&&N_1=n_1r_1^{t_0}, \\ \nonumber
&&N_2=n_2r_2^{t_0},
\end{eqnarray}
where the definition of $N$, $N'$, $N_1$ and $N_2$ are same as $n$, $n'$, $w$, $w'$, $n_1$ and $n_2$ respectively but with new initial condition. We can get 
\begin{eqnarray}
\ln W'=\frac{\ln r_2}{\ln r_1}\ln W+\ln (N_2w_2)-\frac{\ln r_2}{\ln r_1}\ln (N_1w_1) \nonumber
\end{eqnarray}
from the first four equations and the last two equations make this equation be equivalent to Eq.(10).
\\

\bibliographystyle{unsrt}
\bibliography{mybibfile}

\end{document}